\documentclass[aps,prl,twocolumn,groupedaddress,showpacs,showkeys]{revtex4}
\usepackage{epsfig}

\def\dd{{\rm d}}
 
\begin{document}

\title{From Dimensional to Cut-Off Regularization
\footnote{Supported by the Forschungszentrum FZ J\"ulich (COSY)}}

\author{M. Dillig\footnote{mdillig@theorie3.physik.uni-erlangen.de}} 
\affiliation{Institute for Theoretical Physics III\footnote{preprint FAU-TP3-06/Nr. 07} \\
University of Erlangen-N\"urnberg, Staudtstr.7, D-91058 Erlangen, Germany }


\begin{abstract}

We extent the standard approach of dimensional regularization of Feynman
diagrams: we replace the transition to lower dimensions by a 'natural'
cut-off regulator. Introducing an external regulator of mass $(\lambda)^{2 \epsilon}$,
we regain in the limit $\epsilon \rightarrow 0$ and $\epsilon \ge 0$ the results of dimensional and cut-off regularization, respectively. We demonstrate the
versatility and adequacy of the different regularization schemes for practical examples (such as non covariant regularization, the axial anomaly or regularization in effective field theories).

\end{abstract}

\maketitle

Renormalization is a genuine feature of relativistic field theories: in the
presence of the nontrivial physical vacuum all observables result from
the renormalization of unobservable `bare' properties, even in finite theories
\cite{Kato}, \cite{ZJ}. Among different schemes \cite{Collins}, dimensional regularization (DR)
\cite{tHooft}, \cite{Bollini}, \cite{Ashmore}, \cite{Cicuta} of Feynman diagrams
is rather popular:
 both its technical simplicity and the preservation
of fundamental symmetries, like gauge invariance and Ward identities, 
renders it the most widely used scheme for practical
calculations (a collection of the basic
 formulae is found in refs. \cite{tHooft}, \cite{Veltman}, 
\cite{Bollininew}). However, with recent developments, such as the chiral perturbation theory, the issue of the proper regularization of effective field theories has gained new interest \cite{Young}, \cite{Meissner}.

\bigskip

The basic idea of dimensional regularization is to
regularize Feynman integrals of arbitrary dimension d by lowering the dimensionality 
to $D=d-2\epsilon$ (with $\epsilon > 0$)
and to isolate infinities as singularities in the analytic continuation $\epsilon \rightarrow 0$; a regulator mass $(\mu^2)^{\epsilon}$ enters from the appropriate
dimensional mass dependence of the diagram. 
Just to remind of the standard relations: a 
simple example
\begin{equation}
I_{DR}(m^2,\mu^2) \,  = \, (\mu^2)^\epsilon \, \int \dd^{d-2\epsilon}k
 \frac{(k^2)^l}{(k^2+m^2)^n} \,,
\label{eq1}
\end{equation}
yields explicitly
%\begin{eqnarray}
%\lefteqn{I_{DR}(m^2,\mu^2)}
%\nonumber\\
%&  =  &
%\pi^{2-\epsilon} \, \frac{(-1)^n}{(2-n)!}
%\left\{\left(\frac{1}{\epsilon}+ 
%\Psi(3-n)\right)(m^2)^{2-n}\right.
%\nonumber\\
%& & \left.-  \, (m^2)^{2-n} \ln \frac{m^2}{\mu^2}\right\} \, .
%\end{eqnarray}

\begin{eqnarray}
I_{DR}(m^2,\mu^2)=\pi^{d/2-\epsilon}
\nonumber \\ 
\left(\frac{\Gamma(l+\frac{d}{2}-\epsilon)}{\Gamma(\frac{d}{2}-\epsilon)}
\frac{\Gamma(n+\epsilon-l-\frac{d}{2})}{\Gamma(n)}(\frac{\mu^2}{m^2})^{\epsilon}\right) (m^2)^{\frac{d}{2}+l-n} \,.
%(\frac{\Gamma(l+\frac{d}{2}-\epsilon)}{\Gamma(\frac{d}{2}-\epsilon)}
%\frac{\Gamma(n+\epsilon-l-\frac{d}{2})}{\Gamma(n)}(\frac{\mu^2}{m^2})^{\epsilon%}(m^2)^{\frac{d}{2}+l-n}) \,.
\end{eqnarray}

\bigskip

In this note we would like to extent and explore DR, such as for treating the axial anomaly \cite{Adler}, \cite{Jackiw}, its applicability to non covariant Feynman diagrams \cite{LeibbrandtW}, \cite{Holstein}, or for a direct relation to cut-off regularization \cite{Pauli}
and the underlying physical scales, a regularization scheme, as frequently favored in phenomenological effective field
theories.

\vskip 0.6cm

Our main goal in this note is to extent dimensional regularization, calling it natural regularization (NR),
which both preserves the simplicity and the technical formalism of DR and establishes 
simultaneously a direct link to cut-off regularization (CR). To motivate our
generalization,
we formally transform a regulated $d-2\epsilon$ dimensional Feynman integral 
back to the d-dimensional space upon introducing the new (dimensionless) variable 
\begin{equation}
\frac{k^2}{m^2} \, = \, x^{2+\epsilon} \; ; \; \; \,  
\int \dd^{d-2\epsilon} \left(\frac{k}{m}\right)
\, = \, \int \dd^d x \, ,
\end{equation}
which preserves, upon introducing the regulator mass $(\mu ^2)^{\epsilon}$ 
the results obtained above in DR.
The direct relation to
 cut-off regularization is now established
in modifying eq. (3) by  a cut-off mass M such, that in the limit of M being equal to the mass
scale m of the integral itself, i. e. $M \, = \, m$ 
DR is recovered, and at the same time provides a direct route to CR. Explicitly,
we introduce a rather general regulating factor as
\begin{equation}
\int \dd^d k \, f(k^2,m^2) \; \rightarrow \;
\int \dd^d k \, f(k^2,m^2) \;
\mbox{\boldmath
$\left(\frac{\lambda^2}{k^2+M^2}\right)$\unboldmath}^\epsilon \,,
\label{eq5}
\end{equation}
where the scales $\lambda$ and M are identified with the external
scale parameters in the problem.
Evidently, the extension above provides flexibility enough to combine 
dimensional and cut-off regularization: choosing in the
previous equation
{\small
\begin{eqnarray}
\lefteqn{\lambda^2 = \mu^2, M^2=m^2, \; 
\mbox{ with } \epsilon \rightarrow 0:} 
\nonumber\\
& & I^{\epsilon}{(k^2,m^2,\mu^2)}\; = \; (\mu^2)^{\epsilon}
\int \dd^d k \, \frac{(k^2)^l}{(k^2+m^2)^{n+\epsilon}} \, ,
\nonumber\\
\label{eq6}
\end{eqnarray}
}yields the correct expression for natural or, equivalently,
dimensional regularization, whereas \\
{\small
\begin{eqnarray}
\lefteqn{\lambda^2 = \, M^2 \,= \,\Lambda^2, \;
\mbox{ with } \epsilon > 0:}
\nonumber\\
I^{\epsilon}{(k^2,m^2,\Lambda^2)}
 & = & \int \dd^d k \frac{(k^2)^l}{(k^2+m^2)^n} \,
\Big(\frac{\Lambda^2}{k^2+\Lambda^2}\Big)^{\epsilon} \, 
%\left I^{\epsilon}{(k^2,m^2,\Lambda^2)}
% & = & \int \dd^d k \left\frac{(k^2)^l}{(k^2+m^2)^n}\right \,
%\left(\frac{\Lambda^2}{k^2+\Lambda^2}\right)^{\epsilon} \, ,
\end{eqnarray}
}yields the standard representation for cut-off regularization
(the normalization point for the cut-off can be chosen at will
by replacing $\Lambda^2 \rightarrow (\Lambda ^{'})^2$ in the numerator
of the regulator). The equivalence of the generalized formulae above to
 DR and CR is
easy to show using appropriate integral representations \cite{Gradstin}:
the final result agrees with both limits, except of trivial constants,
which may be removed in the subtraction scheme.

\bigskip

Deferring further details of our formalism to a forthcoming note
(a comprehensive list of references is given there), we just collect the 
most important consequences and characteristics of the extension above :

\begin{itemize}

\item
{\bf Regularization of one-loop integrals in NR}:

Coming back to eq. (1) we obtain in NR

\begin{eqnarray}
I_{NR}(m^2,\mu^2)=\pi^{d/2}
\frac{\Gamma(l+\frac{d}{2})}{\Gamma(\frac{d}{2})}\; 
\frac{\Gamma(n+\epsilon-l-\frac{d}{2})}{\Gamma(n+\epsilon)} \
\nonumber\\
(\frac{\mu^2}{m^2})^{\epsilon}
(m^2)^{\frac{d}{2}+l-n} \, ,
\end{eqnarray}
which differs from DR in eq. (2) only in irrelevant constants;

\item
{\bf Consistency of $\int \dd^d k \, (k^2)^n \; = \; 0$}:

Regularizing naturally
\begin{equation}
\int \dd^d k \, (k^2)^n  \, \left(\frac {\mu^2}{k^2+m^2}\right)^{\epsilon} \;, 
\end{equation}
yields, upon identifying the arbitrary mass scale m with the only
scale in the integral, the regulator mass, i. e. $m^2= \mu ^2$, a
constant piece, which is removed in the subtraction scheme;

\item

{\bf Multi-loop integrals} \cite{Ramondbook}:

Again we provide only one classical example, i. e. the `setting sun'
diagram (in 4 dimensions) {\small
\begin{eqnarray}
%\lefteqn{I(m^2,p^2)=}
I(m^2,p^2)=
\int\int\dd k \dd q
\frac{1}{(k^2+m^2)(q^2+m^2)^2((q-k+p)^2+m^2)} \, ,
\end{eqnarray}
}which immediately yields after Feynman
parametrization and  upon performing the regularized integrals
in momentum space
\begin{widetext}
\begin{equation}
I(m^2,p^2) \, = \,
(\pi^2)^2 \mu^2 \frac{\Gamma(\epsilon)}{\Gamma(2+\epsilon)}
\int _{0}^{1} \dd\alpha \int _{0}^{1} \dd\beta\;
\frac{\beta^{\epsilon -1}(1-\beta)}{
\left\{\alpha(1-\alpha)\beta (1-\beta)p^2
+ m^2 \left[ \alpha(1-\alpha)\beta
+(1-\beta)\right]\right\}^{\epsilon}} \, .
\end{equation}
\end{widetext}
The result agrees with DR: the double singularity is extracted from
$\Gamma(\epsilon)$ and from the integration (by parts) of the parameter
$\beta$;

\item
{\bf Non covariant (split) regularization} \cite{LeibbrandtW}, \cite{Leibbrandt}:

NR allows to introduce an arbitrary number of splittings
(regulators and mass scales) as
\begin{eqnarray}
\lefteqn{(\frac {\mu^2}{k^2+m^2})^{\epsilon}}
\nonumber\\
& \rightarrow &
(\frac {\mu_1^2}{k^2+m_1^2})^{\epsilon-\epsilon_1} \; 
(\frac {\mu_2^2}{k^2+m_2^2})^{\epsilon_1 - \epsilon_2} \; ... ; 
\end{eqnarray}
and thus a natural application to non covariant formulations, such as gauge
theories or
to regularization on the light cone,
by either introducing various infinitesimal
 external parameters or by regularizing arbitrarily one
 integration variable with just one single infinitesimal 
parameter either before or (for factorized integrals) after introducing
Feynman parametrization;

\item
{\bf Gauge invariance} \cite{tHooft}: 

to recover gauge invariance (which is lost in going over from DR to NR
by the replacement $d-2 \epsilon \rightarrow d$), we remind that we have
two scales in any integral
\begin{eqnarray}
\lefteqn{\int \dd^d k  \, \frac {(k^2)^l}{(k^2+m^2)^n} \rightarrow}
\nonumber \\
& & \int \dd^d k   \, \frac {(k^2)^l}{(k^2+m^2)^n} \,
\left(\frac {\mu^2}{k^2+m^2}\right)^{\epsilon}
\nonumber\\
& & \leftrightarrow 
\int \dd^d k  \, \left(\frac {\mu^2}{k^2}\right)^{\epsilon}
\; \frac {(k^2)^l}{(k^2+m^2)^n} \, ,
\end{eqnarray}
which recovers gauge invariance upon an appropriate combination of the
two regularization modes;

\item
{\bf axial anomaly} \cite{ZJ}, \cite{Collins}, \cite{Bardeen}, \cite{Jeg} :

\noindent
%The 
%simultaneous Ward identities cannot be restored by an appropriate
%shift of the integration variable, as the integrals are linearly divergent
%in 4-dimensional space.
As well known, 'standard' dimensional regularization faces problems for the axial anomaly, as
the definition of the pseudo-scalar matrix $\gamma_5 \, = \,i 
\epsilon_{\alpha \beta \gamma \delta} \gamma_{\alpha} \gamma_{\beta}
 \gamma_{\gamma}
\gamma_{\delta}$ via the antisymmetric
$\epsilon$-tensor is genuinely related to 4 dimensions; 
technically, the problem is overcome by formal 
extensions of $\gamma_5$ to $d \, > \, 4$, defining
appropriate commutation
and anti-commutation relations with the matrices $\gamma_\mu$.
%\cite{tHooft}, \cite{Bardeen}, \cite{Breiten}, \cite{Chanowitz}).\\

NR provides a consistent regularization: upon Feynman parametrizing and regularizing the axial
triangle yields
\begin{widetext}
\begin{equation} 
T_{\mu \nu \lambda} \, = \,
-i \int \dd\alpha \int \dd\beta \int \dd^4k
\frac { (\mu^2)^\epsilon \;
{\rm Tr}\left[(\slash{\!\!\!k}+ \slash{\!\!\!k_1}+m)\,\gamma_\mu \, 
(\slash{\!\!\!k}+m) \,
\gamma_\nu \,
(\slash{\!\!\!k}- \slash{\!\!\!k_2}+m) \, \gamma_\lambda \gamma_5\right]}{\left\{
\alpha\left[(k+k_1)^2-m^2\right]+
\beta (k^2-m^2)+(1-\alpha-\beta)\left[(k-k_2)^2-m^2\right]
\right\}^{3+\epsilon}}\,.
\end{equation}
\end{widetext}
The convergent loop integral is now readily performed after an
 appropriate shift in the loop momentum;
%extracting
%simultaneously the $\epsilon$-tensor piece without destroying the
%nature of the $\gamma_5$;

\item

{\bf Optimal subtraction scheme}:

From the results above,
we realize that NR provides much flexibility for different 
regularization procedures by an appropriate superposition of the
different prescriptions, such as
{\small
\begin{equation}
I(m^2) \rightarrow
 \lambda \,\int \dd^4 k \, \frac{k^2}{(k^2+m^2)^{1+\epsilon}}
 +  (1-\lambda) \, \int \dd^4 k \,
\frac{(k^2)^{1+\epsilon}}{k^2+m^2} \,,
\end{equation}
} with an arbitrary constant $\lambda$. This
flexibility in the regularization might be either used to improve the
convergence of a perturbative expansion or, alternatively, these additional
constants have no physical content and thus can be completely
removed (along the $\overline {MS}$
subtraction scheme \cite{ZJ});

\item

{\bf Integrating out effective field theories}:

Already a simple example exhibits some shortcomings of DR
in context with effective field theories: the well
defined integral in dimensional regularization
\begin{equation}
I_{DR}(m,\epsilon) =
(\mu^2)^\epsilon \int \dd^{4-2\epsilon}k\, \frac{1}{(k^2+m^2)}
 \; \sim m^2 \, \ln (\frac{\mu^2}{m^2}) \,,
\end{equation}
vanishes identically according to eq. (8), if expanded around the mass parameter m
\begin{equation}
I_{DR}(m,\epsilon)^{EFT}
\; = \; \sum^\infty_{n=0}
\frac{(-1)^n}{(m^2)^{n+1}}
 \int \dd^{4-2\epsilon} (k^2)^n \; = 0 \; .
\end{equation}
Natural regularization yields a finite result, upon expanding
the denominator and regularizing it with the regulator mass $\mu$. \\

As an extension of this example we integrate out
masses in effective field theories. Explicitly we find for
\begin{eqnarray}
I(m^2,M^2) & = & \int \dd^4 k \frac{1}{(k^2+m^2)} \frac{1}{(k^2+M^2)}
\nonumber\\
& = & \pi^2 \frac{\Gamma(-1+\epsilon)}{\Gamma(1+\epsilon)} \;
\frac{1}{(M^2-m^2)}
\\
&& \kern4em\left(m^2 \ln \frac{m^2}{\mu^2} \, - \, 
M^2 \ln \frac{M^2}{\mu^2}\right) \, .
\nonumber
\end{eqnarray}
This standard result is recovered in NR, upon expanding the
denominators symmetrically around M and m 
and after regularizing via an effective regulator scale,
\begin{eqnarray}
\lefteqn{I(m^2,M^2)}
\nonumber\\
& = & \int \dd^4 k \frac{1}{(k^2+m^2)} \frac{1}{k^2+M^2}
\left(\frac{\mu^2}{k^2+M^2}\right)^{\epsilon}
\nonumber\\
& = & - \sum \left(-\frac{1}{M^2}\right)^{n+1} \, \int \dd^4 k 
\frac{(k^2)^n}{(k^2+m^2)} \left(\frac{\mu^2}{k^2+m^2}\right)^{\epsilon}
\nonumber\\
& = & \pi^2 \, \frac{\Gamma(-1+\epsilon)}{\Gamma(1+\epsilon)} \,
\frac{m^2}{(M^2-m^2)} \, \ln \frac{m^2}{\mu^2} \, + \, (M \leftrightarrow m)  \,
;
\nonumber\\
\end{eqnarray}

\item
{\bf Unifying DR and CR}:

\bigskip

NR provides a transparent unification of both schemes: upon introducing
2 independent
external parameters $\lambda$,M for the regularization, we obtain
for a characteristic example (compare eq. (4))
\begin{eqnarray}
\lefteqn{J(\lambda,M, \epsilon) \, = \, \int \dd^3k\,
\frac{k^2}{k^2+m^2}\left(\frac{\lambda^2}{k^2+M^2}\right)^\epsilon}
\nonumber\\
& = & \pi^{3/2}
\frac{3}{2} \frac{\Gamma\left(-\frac{3}{2}+
\epsilon\right)}{\epsilon \,\Gamma(\epsilon)} 
\left(\frac{\lambda^2}{m^2}\right)^{\epsilon}\, 
m^3
\nonumber\\
& & \kern2em F\left(\epsilon-\frac32,\epsilon,\epsilon+1,
- \frac{M^2-m^2}{m^2}\right),
\end{eqnarray}
(with the hypergeometric function $F(a,b,c,z)$), which holds for all values
of $\epsilon$. Evaluating F, we immediately obtain, for example, in

\begin{eqnarray}
\nonumber
DR: \; \lambda = \mu; M = m \; : \: \epsilon \rightarrow 0;\\
\nonumber
CR \, (Dipole): \; \lambda = M = \Lambda \; : \: \epsilon = 2;\\
\end{eqnarray}
furthermore, similarly as in split regularization, we may introduce in CR
different discrete or continuous external cut-off scales both in $\Lambda$ 
and $\epsilon$,
such as 
\begin{equation}
 \sim \int J(\Lambda, \epsilon) \, \rho (\Lambda, \epsilon) \, d \Lambda 
d \epsilon
\end{equation}
(with the density distribution $\rho(\Lambda, \epsilon)$ appropriately normalized) ;

\item
{\bf Functional relation of a cut-off and regulator mass in CR, DR}:

Though a rigorous one-to-one comparison explicitly depends on the diagrams under consideration, the global relation is readily extracted for the two most frequent examples, isolating a quadratic
 or logarithmic divergence in NR in 4 (in general: d) dimensional Euclidean space, upon
evaluating the combination of integrals 

\begin{equation}
I_1^{\epsilon}(m^2,\mu^2)\, - m^2 \, I_2^{\epsilon}(m^2,\mu^2)
\end{equation}
for
\begin{equation}
I_n^{\epsilon}(m^2,\mu^2) \, = \,
\int \dd^4 k \frac{\mu^2}{(k^2+m^2)^{n+\epsilon}}
\end{equation}
both in dimensional and cut-off regularization. Comparing the limit
 $\epsilon \rightarrow 0$, with cut-off regularization
for a scale mass $\Lambda$ yields 
\begin{equation}
\frac{\Lambda^4}{\Lambda^2+m^2} \sim \Lambda^2 \; = \; - \frac{2}{\epsilon} \,
\left(\frac{\mu^2}{m^2}\right)^{\epsilon} \, m^2 \; .
\end{equation}
for $\Lambda \rightarrow \infty$ (and $\epsilon \rightarrow 0$);

\item

{\bf Sharp momentum cut-off versus soft form factor}: 
 
For integrals with a vanishing mass scale, DR according to eq. (8) is in
apparent contrast to CR, yielding for a sharp cut-off $\Lambda$ 
\begin{equation}
I_{C}^{(n)}(\Lambda) \; = \;
\int_{0}^{\Lambda}  \dd^4 k (k^2)^n \; = \; \pi^2 \, 
\frac{(\Lambda^2)^{n+2}}{n+2} \; ,
\end{equation}

In NR we obtain the same result, upon introducing a regulator mass
 $\mu$ 
for a finite $\epsilon$, i. e.
\begin{eqnarray}
I_{NR}^{(n)} & = & \int \dd^4 k (k^2)^n\,
\frac{(\mu^2)^{\epsilon}}{(k^2+\mu^2)^{\epsilon}}
\\
& = & \pi^2 \,
\frac{\Gamma(n+2)}{\Gamma(2)}
\frac{\Gamma(\epsilon-(n+2))}{\Gamma(\epsilon)}
(\mu^2)^{n+2} \; ,
\nonumber
\end{eqnarray}

which evidently agrees with cut-off regularization upon identifying
$\Lambda = \mu$ for $\epsilon \, = \, n+3$ (reflecting the asymptotic behavior of the integral like a monopole $\sim 1/k^2$). Clearly, for effective field theories 
this form of regularization is appealing, as it explicitly reflects the typical scale of the underlying
physics, i. e. inverse size of the effective degrees of freedom.

The example above holds also for cut-offs via soft form factors: a quantitative relation between a sharp and a soft cut-off is readily 
established such as for
\begin{equation}
I_{\rm SC}(m^2,\Lambda^2) \; = \;
\int_0^\Lambda \dd^4 k\,\frac{(k^2)^\ell}{(k^2+m^2)^n} \; ,
\end{equation}
upon introducing the transformations
\begin{equation}
q^2 = \frac{\Lambda^2}{\Lambda^2-k^2}\,k^2 \;\longleftrightarrow\;
k^2=\frac{\Lambda^2}{\Lambda^2+q^2}\cdot q^2\,,
% {\bf formula (5) \; ,}
\end{equation}
ending up with
\begin{widetext}
\begin{equation}
I_{\rm SC}(m^2,\Lambda^2)\,=\,
\frac{(\Lambda^2)^\ell}{(\Lambda^2+m^2)^n}
\int\dd^4 q\left(\frac{\Lambda^2}{\Lambda^2+q^2}\right)^3
\left(\frac{q^2}{q^2+\Lambda^2}\right)^\ell
\left(\frac{q^2+\Lambda^2}{q^2+
\frac{\Lambda^2}{\Lambda^2+m^2}\,m^2}\right)^n.
% {\bf formula (6)\; .}
\end{equation}
\end{widetext}
The result clearly demonstrates the equivalence of sharp and soft cut-offs:
independent of the explicit momentum dependence of the integral
under investigation, sharp cut-off regularization in 4 (d in general) dimensions
is strictly equivalent to a soft cubic form-factor, guaranteeing a
strict convergence of the integral considered;

\item

{\bf Incorporating external scales in effective field theories}
 \cite{Young}, \cite{Meissner},
\cite{Chiral}, \cite{Lehmann}, 
\cite{Borasoy}:

Effective field theories,
such as Chiral Perturbation 
($\chi$PT) 
 or Heavy Baryon Effective Theory (HBETh)
 are, as expansions in powers of the internal momentum scale,
in the strict sense non-renormalizable field theories; as a consequence
 the transition to effective  
objects introduces typical scales of their spatial extension. 
The origin of the problem is the very nature of
diverging diagrams in NR (DR), yielding just from power counting (compare eqs. (1,2))
\begin{equation}
I^{nl}_{NR} \; = \; \int \dd^{d} k \,  \frac{(k^2)^l}{(k^2+m^2)^n} \, \sim
 \, (m^2)^{\frac{d}{2}+l-n} ln(\frac{m^2}{\mu^2})\, .
\end{equation}
Evidently, the only physical scale in NR (DR) is the mass m of the intermediate particle, which sets the scale just for any momentum dependence 
of the integrals considered. \\

Already a very simple example strikingly exhibits the consequences in practical calculations: comparing the self energy correction of, say, a baryon, induced on the one-loop level by the exchange of mesons with masses
 m, m', respectively. In DR their contribution scales qualitatively with the ratio 
\begin{equation}
 \frac {\Sigma_{m'}}{\Sigma_{m}} \sim (\frac{m'}{m})^{\frac{d}{2}+l-n} \, ,
\end{equation}
Thus for $m^{'}>>m$ and for the momentum expansion in effective
 field theories with typically $\frac{d}{2}+l >> n$, in a perturbative
loop expansion heavy meson exchange (with mass m') dominates. This contradicts  an intuitive physical understanding, for example from the uncertainty principle, from which $\Sigma^{'} << \Sigma$ is expected. Only introducing for example external 
 form factors garantees the
right scales of the corresponding self energies.

\end{itemize}

We summarize briefly the main findings in our short note.
In NR, our starting point
is --- compared to dimensional regularization --- slightly different:
while DR enforces convergence of diverging integrals by dimensionally reducing the 
corresponding phase space, NR cuts down the high momentum components by a dimensionless regulator strictly in
4 (in general: in $d$) dimensions. Without loosing the technical simplicity
 of DR, this simple modification introduces the regulator 
scale in a very natural way: the
regularization itself allows a systematic and well-defined finite extension to singular
diagrams of arbitrary order, by providing 
a direct route from DR to
cut-off regularization.
While DR leaves the internal mass scale $m$ of integration in
momentum space unchanged, NR allows for the introduction of  external mass scales, 
for example via conventional monopole or dipole form factors,
 as characteristic for phenomenological low energy nuclear physics (for further details and
quantitative insight into these and related questions we refer to
a comprehensive forthcoming article).

\vskip 0.5cm

We like to thank A. W. Thomas for discussions on the regularization of effective field theories.

%\bibliography{Nat}

\end{document}